\documentclass[aps, prb, twocolumn]{revtex4}
\usepackage{ams}
\usepackage{graphicx}
\usepackage{color}

\begin{document}

\title{Binding energy of negative trions at high magnetic fields in a CdTe quantum well}
\date{\today}
\author{Pawe{\l} Redli\'{n}ski}
\email{pawel.redlinski.1@nd.edu} \affiliation{Department of Physics,
University of Notre Dame, Notre Dame, Indiana 46556}

\begin{abstract}
We present results of numerical calculations of electronic states of
an exciton and a trion ($X^-$) in a CdTe quantum well at magnetic
fields $B$ up to 150~T. The exciton state and the trion states were
computed using a variational procedure. We estimated the binding
energy of $X^-$ in a singlet state with the z-component of angular
momentum $L_z=0$, as well as the binding energy of the $X^-$ in a
triplet state with $L_z=-\hbar$. Recent experimental results show
that even up to 44~T the binding energy of negative trion in a
singlet state is bigger than in a triplet state. We show that at a
critical field $B_c\approx65$~T the two binding energies cross and
above $B_c$ the binding energy of $X^-$ in a triplet state is bigger
than in the singlet state.
\end{abstract}



\maketitle



After the discovery of quantum mechanics at the beginning of the
20th century, the first to be studied to support new theory were
light atoms and ions such as H, H$^-$, H$_2$ and He. Although only
the non relativistic Schr\"{o}dinger equation of the hydrogen atom
can be solved analytically, these few-particle quantum systems can
still be satisfactorily described using numerical methods. Analogous
complexes can be found in semiconductor physics. An electron in the
conduction band (which we will simply refer to as "an electron") is
similar to the free electron, and the valence band hole (carrying
positive charge) is an analog of the proton. By binding an electron
and a hole we form exciton ($X$) which is the analog of the hydrogen
atom. Adding a second electron (or a second hole) to the exciton we
then form a negatively (positively) \textit{charged exciton}. The
names negative ($X^-$) or positive ($X^+$) \textit{trion} are also
used as a alternative. These three-particle complexes are analogs of
the $H^-$ ion (the importance of which was first noted by 1983 Nobel
prize laureate S. Chandrasekhar) and the $H_2^+$ ion, respectively.
By combining two excitons one can then form the bi-exciton state,
which is a close analog of the H$_2$ molecule. As in the case of the
early universe, when light elements were formed first, in the last
100 years we are again discovering and studing 'light elements' in
semiconductor physics\cite{Lampert}.

Let us take a closer look at the similarities and differences
between the negatively charge exciton and the H$^-$ ion. The main
difference is that the characteristic energy scale (the effective
Rydberg $Ry^*$) and the characteristic length scale (the effective
Bohr radius $a_B^*$) are much smaller in semiconductor than in
atomic physics. Typically, in a bulk CdTe $Ry^*\approx 10$~meV and
$a_B^*\approx 70$~\AA, first, this is because the effective masses
of the electron and of the hole are smaller in a semiconductor than
in the vacuum. And second, the dielectric constant $\epsilon$ of a
typical semiconductor (e.g. like CdTe) is around 10 while in atomic
physics $\epsilon=1$. This difference in the characteristic scale
allows electric\cite{Dacal} and magnetic field-related phenomena to
be studied in the laboratory using much lower values of the
corresponding fields. For example, at $B=100$~T the cyclotron energy
of a particle with the effective mass of $0.5\;m_0$ is
$\hbar\omega_c \approx 11.5 \text{meV} \approx Ry^*$, whereas in
atomic physics the cyclotron energy is much smaller than the Rydberg
energy.

An additional advantage of using a semiconductor to study
multi-particle complexes is the possibility of decreasing the
dimensionality of the problem. Using semiconductor quantum
wells\cite{Whittaker} (QWs), quantum wires\cite{Otterburg} and
quantum dots\cite{Xie}, we are able to study the properties of these
complexes in quasi two-, quasi one- and quasi zero-dimensional
systems, respectively. Decreasing dimensionality allows, for
example, trion states to be experimentally observable, in contrast
to the bulk case, where trion states have never been measured. In
this report we concentrate on a quasi two-dimensional system
consisting of type I quantum wells, see Refs.~\onlinecite{Whittaker,
Redlinski} and references therein.

We will consider a negatively charged exciton composed of two
identical electrons and one hole. Because two identical particles
are involved it is always possible\cite{Mattis} to factor the wave
function into a spin-dependent part and an orbital-dependent part
and at the same time to ascertain that the wave function is totaly
antisymmetric (Pauli exclusion principle). This implies that the
trion can be found in a singlet or in a triplet state of the two
electrons. Second, the orbital part of the trion wave function in a
QW is characterized by the z-component of the total orbital angular
momentum of three particles: $L_z=0,\pm\hbar,\pm2\hbar,...$.
Generally the energy spectrum of the trion is quite complex but in
this paper we will consider only the singlet state, with the
z-component of the angular momentum $L_z=0$; and the triplet state,
with $L_z=-\hbar$.

Recently there appeared two articles\cite{Imanaka, Astakhov}
concerning, among other things, the binding energy of the negatively
charged excitons in CdTe QW structure. In both experiments the
quantum structure was almost identical: in the first group
(Ref.~\onlinecite{Imanaka}) trion states were studied using a QW
with the width $L=100$~\AA~while the second group
(Ref.~\onlinecite{Astakhov}) used a wider quantum well, $L=120$~\AA.
The magnesium content in the barriers given by the authors was
almost the same in both cases: 13.5\% and 15\%, respectively. We
believe that the main difference concerning both structures is
related to the fact that in the first experiment the carrier
concentration was estimated to be 20 times larger than in the second
experiment. Both groups claim that they did not find any evidence of
crossing between the binding energies of the single and the triplet
states up to highest external magnetic fields available to them
(35~T and 44~T, respectively). On the other hand the
photoluminescence spectra of both experiments show crossing between
singlet and triplet states around $B=24$~T.

In this report we show that the binding energy of the triplet state
of $X^-$ ($L_z = - \hbar$) is larger than the binding energy of the
singlet state ($L_z = 0$) at high magnetic fields, above $B_c
\approx 65$~T. The calculation procedure is based on our previous
considerations presented in Ref.~\onlinecite{Redlinski}. In the
calculations we assume the valence band barriers to be $V_h=80$~meV,
and the conduction band barriers to be $V_e=180$~meV, which we
believe correspond to a 15\% Mg content in the barriers. Our quantum
well width is fixed at $L=120$~\AA, and we use the following
materials parameters: the mass of the hole (heavy hole) in the
growth (001) direction $m_{hz} = 0.48\;m_0$; the mass of the heavy
hole in the plane of the QW $m_{h\rho} = 0.37\;m_0$; the electron
mass $m_e=0.096\;m_0$; and dielectric constant $\epsilon$~=~10.4.
These parameters are assumed to be the same in the QW and in the
barriers.

\begin{figure}
\includegraphics[height=.27\textheight]{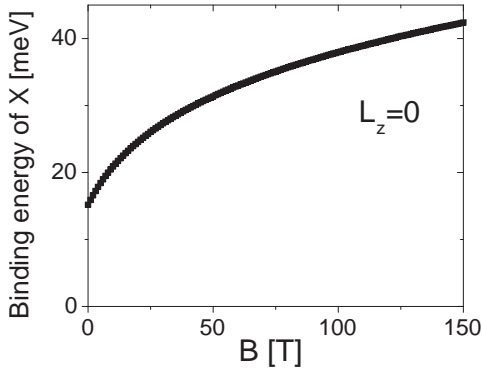}
\caption{Binding energy of the exciton with angular momentum $L_z$=0
as a function of magnetic field $B$. The binding energy depends
primarily on the in-plane heavy hole mass.}\label{figure1}
\end{figure}
Because we are considering binding energies of the exciton and of
the  trion, we can ignore the Zeeman part of the interaction between
the spins of the particles in an external magnetic
field\cite{Redlinski,Vanhoucke}. Schematically the Hamiltonian of
the exciton (N=2) or the trion (N=3) system can be written as
\begin{eqnarray}\label{Ham}
H(N) & = & \sum_{i=1}^N T_i(B)    +   \sum_{\begin{array}{c}
                            i,j=1 \\
                            i\ne j \\
                          \end{array}}^N
                          V^{Coul}_{i,j}  \\\nonumber
                          & = & H_0(N)    +   H_{int}(N).
\end{eqnarray}
Here $H_0(N)$ is the sum of the kinetic energies $T_i(B)$ of the
$i^{th}$ particle in an external magnetic field $B$, and
$H_{int}(N)$ is the sum of the Coulomb interaction energies
$V^{Coul}_{i,j}$ between particles $i$ and $j$.


In order to calculate the binding energy of the trion, we must first
calculate the binding energy of the exciton. The calculation
involving $X$ are much less complicated than those involving $X^-$,
because of the larger number of degrees of freedom in the case of
the trion. In Fig.~\ref{figure1} we show the dependence of the
binding energy of $X$ up to 150~T. Only the optically-active $X$
state with the z-component of the angular momentum $L_z=0$ is shown.
The calculations reproduces well known functional form dependence
($\sim \sqrt{B}$) of this state as a function of the magnetic
filed\cite{Blinowski}. At $B=0$ the binding energy is about 15~meV,
i. e., $1.5\;Ry^*$.

\begin{figure}
\includegraphics[height=.29\textheight]{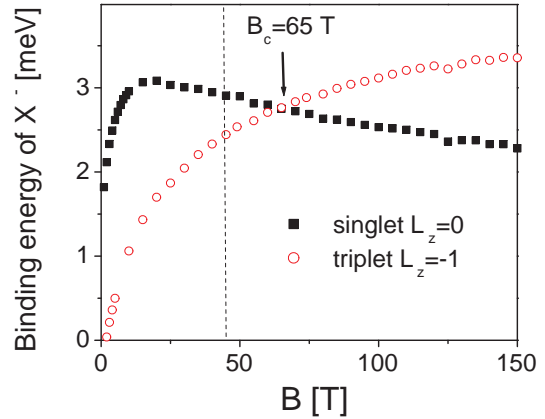}
\caption{Binding energy of $X^-$ as a function of magnetic field
$B$. The calculation is made for the singlet state with $L_z$=0 and
for the triplet state with $L_z=-\hbar$. At $B=0$ the triplet state
is not bound. At $B_c\approx 65$~T the binding energies of the
singlet and triplet states cross. The vertical dashed line indicates
the largest magnetic field (44~T, Ref.~\onlinecite{Astakhov}) which
had been experimentally accessible to study the $X^-$ complexes in a
CdTe quantum well.}\label{figure2}
\end{figure}

In Fig.~\ref{figure2} we show the binding energy of the $X^-$ in
singlet and triplet configurations with $L_z=0$ and $L_z=-\hbar$,
respectively.
We see immediately that at $B = 0$~T the only bound state is a
singlet state. In our notation this means that the binding energy of
the triplet state is negative. The crossing between the singlet and
the triplet states appears at the crossing field $B_c = 65$~T. Such
a crossing is expected at some magnetic field $B=B_c$ because, as
was discussed by Whittaker and Shields in
Ref.~\onlinecite{Whittaker}, at very high magnetic field the triplet
state is the ground state of the $X^-$. The question remains:
\textit{What is the value of the crossing field $B_c$?}. The value
of $B_c$ is substantially smaller as compared to the corresponding
value in the case of the $H^-$ ion, where $B_c$ of thousands of
Tesla is expected. Looking at Fig.~\ref{figure2}, the slopes of both
curves are relatively small. This feature is a disadvantage from our
point of view, because it means that small changes in the slopes of
the curves can substantially change their crossing point $B_c$. Our
results can be directly compared to results of the experiment done
by Astakhov \textit{et al}. In Fig.~3c the Authors plotted binding
energy of trion in singlet state with $L_z=0$ (in their notation
$T_{s}$) and in triplet state with $L_z=-\hbar$ (in their notation
$T_{td}$). Both energies do not cross up to B=44~T. We want to
stress that the \emph{photoluminescence spectra} of both experiments
show crossing between singlet and triplet states around 24~T but
photoluminescence energy is \emph{not} the same as binding energy.

In order to test our approach, we have applied the following
procedure. In our algorithm we removed the Coulomb interaction term
$H_{int}(N)$ from Eq.~(\ref{Ham}) and then applied our
minimalization procedure. This allowed us to compare numerical
results with exact values. Table~\ref{table1} show such a comparison
for the case of $X^-$. It is seen that at up to 150~T the difference
between our numerical method and the exact result is not greater
than 0.1~meV. Additionally, the relative error even at $B=150$~T is
only $3*10^{-4}$. A similar analysis was made for the case of the
exciton, where we obtained even better results: at $B=150$~T
difference between the exact value and our results of minimalization
is only $10^{-4}$~meV.
\begin{table}
\caption{\label{table1} The ground state energy of a free
three-particle system calculated exactly and calculated using our
approach at $B$=50, 100 and 150~T. The last row indicates the
difference between the two procedures. Even at $B$=150~T our
approach gives very good agreement with the exact value.}
\begin{ruledtabular}
\begin{tabular}{cccc}
B [T] & 50 & 100 & 150\\
\hline
exact value [meV]   &   68.12   &  136.24   & 204.35\\
our estimation [meV]&   68.14  & 136.28  & 204.42\\
\hline
difference [meV]    &   0.02 & 0.04 & 0.07\\
\end{tabular}
\end{ruledtabular}
\end{table}
This procedure is the first of two testing procedures which we
applied to our calculations concerning the problem of the trion. The
second procedure was to show the convergence of the results when the
number of base functions increases\cite{Whittaker}, similar to that
already used in our previous paper\cite{Redlinski}.


In conclusion, our numerical procedure of calculating magnetic field
at which binding energies of the $X^-$ in singlet and in triplet
states are crossing was applied to the CdTe/CdMgTe quantum well
structure. We found that above $B_c=65$~T the binding energy of
$X^-$ triplet state with $L_z=-\hbar$ is larger than the binding
energy of the trion singlet state with $L_z=0$. Up to now there is
no experimental evidence to confirm these findings. We believe
however, that the application of pulsed magnetic field techniques
should make it possible to demonstrate the phenomenon of the
singlet-triplet trion crossing at fields near $B_c=65$~T.

This material is based in part upon work supported by the NSF
DMR02-10519. Any opinions, findings, and conclusions or
recommendations expressed in this material are those of the
author(s) and do not necessarily reflect the views of the NSF.


\end{document}